\documentclass[showpacs,aps,prd,reprint,superscriptaddress,nofootinbib,longbibliography]{revtex4-1}
\usepackage[colorlinks=true, pdfstartview=FitV, linkcolor=magenta,citecolor=blue, urlcolor=magenta,
bookmarks=true, bookmarksnumbered=true, breaklinks]{hyperref}
\usepackage[dvipdfmx]{graphicx}
\usepackage{url}
\usepackage{amsmath,amssymb,bm,color,longtable,mathrsfs,amsfonts,slashed,ulem}

\newcommand{\Slash}[1]{{\ooalign{\hfil/\hfil\crcr$#1$}}}

\newcommand{\beq}{ \begin{eqnarray} }
\newcommand{\eeq}{ \end{eqnarray} }

\begin{document}
\begin{flushright}
\end{flushright}

\title{
Delineating chiral separation effect in two-color dense QCD}

\author{Daiki Suenaga}
\email{suenaga@rcnp.osaka-u.ac.jp}
\affiliation{Research Center for Nuclear Physics, Osaka University, Ibaraki, 567-0048, Japan}

\author{Toru Kojo}
\email{torujj@mail.ccnu.edu.cn}
\affiliation{Key Laboratory of Quark and Lepton Physics (MOE) and Institute of Particle Physics, Central China Normal University, Wuhan 430079, China}

\date{\today}

\begin{abstract}

We study the chiral separation effect (CSE) in two-color and two-flavor QCD (QC$_2$D) to delineate quasiparticle pictures in dense matter from low to high temperatures. 
Both massless and massive quarks are discussed.
We particularly focus on the high density domain where diquarks form a color singlet condensate with the electric charge $1/3$. 
The condensate breaks baryon number and $U(1)_A$ axial symmetry, and induces the electromagnetic Meissner effects. 
Within a quark quasiparticle picture, we compute the chiral separation conductivity at one-loop. We have checked that Nambu-Goldstone modes, which should appear in the improved vertices as required by the Ward-Takahashi identities, do not contribute to the chiral separation conductivity due to their longitudinal natures. 
In the static limit, the destructive interferences in the  particle-hole channel, as in usual Meissner effects, suppress the conductivity (in chiral limit, to $1/3$ of the normal phase's).
This locally breaks the universality of the CSE coefficients, provided quasiparticle pictures are valid in the bulk matter. 
\end{abstract}


\maketitle

\section{Introduction}
\label{sec:Introduction}

The quantum anomaly \cite{Adler:1969gk,Bell:1969ts,AlvarezGaume:1983ig} and topological properties of matter have been attracting much attention in wide area of physics, due to their universal natures valid for abruptly different energy scales  \cite{Vilenkin:1980fu,Kharzeev:2007jp,Fukushima:2008xe,Son:2009tf,Son:2012wh,Stephanov:2012ki,Yamamoto:2015gzz}.
The non-renormalization theorem for the anomaly relation \cite{Adler:1969er,Adler:2004qt,Golkar:2012kb}, as well as the robustness of topology under local perturbations, are the key concepts for such universality.

The chiral conductivities are constrained by such universal relations, especially in the chiral limit (see, e.g. a review \cite{Landsteiner:2016led}). One can also use the relations to constrain or enrich our pictures on the nonperturbative dynamics \cite{Itoyama:1982up,Sannino:2000kg,Gaiotto:2014kfa,Gaiotto:2017yup,Tanizaki:2017mtm,Tanizaki:2018wtg,Wan:2019oax,Tanizaki:2019rbk} which have important implications to the physics of neutron stars (for a review, e.g., \cite{Baym:2017whm}).
With this perspective in mind, in this paper we study the conductivity for chiral separation effect (CSE) \cite{Metlitski:2005pr}, chiral separation (CS) conductivity, in two-color QCD with two-flavor (QC$_2$D) at high baryon density. 
The relation is
\beq
J_i^5 = \bar{q} \gamma_i \gamma_5 q = \frac{\, C \,}{\, 2\pi^2 \,} \mu B_i \,,
\eeq
with $q$ quark fields, $\mu$ quark chemical potential, and $C$ being some constant related to the anomaly.
The dense QC$_2$D matter  \cite{Kogut:2000ek,Kogut:1999iv} can be studied in lattice Monte-Carlo simulations \cite{Boz:2019enj,Iida:2019rah,Astrakhantsev:2020tdl,Boz:2018crd} 
and hence can be used as laboratories to test the concepts and methodologies of dense QCD calculations \cite{Kojo:2021knn,Suenaga:2019jjv,Kojo:2014vja}. The results may be extended to our main target, three-color dense QCD.

The CSE in QC$_2$D was studied for hadronic and quark-gluon-plasma phase \cite{Buividovich:2020gnl}, and was found to agree with theoretical predictions including explicit chiral symmetry breaking.
Meanwhile, for cold dense matter the CSE has not been studied yet. At $\mu\geq m_\pi/2$ ($m_\pi$: pion mass), the dense matter has condensations of diquarks, $\sim ud$, which are color-singlet, $J^P=0^+$, and have the electric charges $1/3$. We call the phase diquark condensed phase.
A number of new ingredients appear here, modifying qualitative pictures conventionally used to explain the CSE \cite{Metlitski:2005pr}; 
(i) A quark acquires the mass gap around the Fermi surface and arguments based on the lowest Landau level no longer directly apply;
(ii) Diquarks are charged and should induce the electromagnetic Meissner effects, expelling magnetic fields from the bulk;
(iii) The spontaneous symmetry breaking demands the improved vertices to fulfill the Ward-Takahashi identity (WTI), and the vertices contain the poles of the Nambu-Goldstone (NG) modes \cite{Nambu:1960tm},
both in the axial vector and the vector vertices.

We will address these qualitative issues within one-loop calculations in the linear response regime, as done for the chiral magnetic conductivity \cite{Kharzeev:2009pj}.
We assume the quasiparticle pictures for quarks with the gap $\Delta$, in a similar manner as in our previous calculations \cite{Kojo:2021knn,Suenaga:2019jjv,Kojo:2014vja}. 
The CS conductivity in the static and dynamic limits will be analyzed for various temperatures.
We found that the dependence on temperatures and the orders of static and dynamic limits can be understood through the particle-hole dynamics;
in the chiral limit and at $T=0$, the static CS conductivity in the diquark condensed (normal) phase acquires $0$ ($2/3$) of the anomaly coefficient from the particle-hole contributions and $1/3$ ($1/3$) from particle-antiparticle contributions.
Thus the static CS conductivity in the diquark condensed phases provides only $1/3$ expected from the universal relation.\footnote{In contrast, the CS conductivity was found to be enhanced by heavy impurities in Refs.~\cite{Araki:2020rok,Suenaga:2020oeu}.}

Inclusion of the NG modes through the improved vertices does not help as they are longitudinal modes and hence decouple.
This suggests that, if the coefficient of the CS conductivity is indeed constrained by the anomaly, 
we should go beyond our quasiparticle descriptions or the linear response regime, e.g., by manifestly treating vortices with fermion zero modes or zero modes on the boundaries of the diquark condensed phase.

This paper is structured as follows.
In Sec.\ref{sec:Model} we set up our frameworks and introduce quasiparticle propagators in the diquark condensed phase.
In Sec.\ref{sec:CSE_conductivity} we summarize the expressions for the CS conductivity within the linear response regime.
In Sec.\ref{sec:Results} we present our numerical results for various chemical potential, temperature, quark mass, and diquark gap. 
In Sec.\ref{sec:Discussions} we address the contributions not included in our quasiparticle calculations in the bulk.
This paper is closed with summary in Sec.\ref{sec:Conclusions}.

\section{Model}
\label{sec:Model}

Here we introduce an effective model to investigate the CSE in QC$_2$D. Previous works by lattice QCD \cite{Boz:2019enj,Iida:2019rah,Astrakhantsev:2020tdl,Boz:2018crd} and model calculations such as the Nambu--Jona-Lasinio 
model \cite{Sun:2007fc,He:2010nb,Andersen:2010vu,Andersen:2015sma}
and the low-energy effective theory based on the nonlinear realization \cite{Kogut:2000ek,Kogut:1999iv}  imply the existence of the diquark condensed phase, or the diquark condensate phase, in a region of $\mu\geq m_\pi/2$ ($\mu$ is a quark chemical potential and $m_\pi$ is a pion mass) at zero temperature in QC$_2$D. The diquark condensed phase is described by $S$-wave, color-singlet and  flavor-singlet diquark condensate $\Delta\sim\langle \psi^T C\gamma_5\sigma^2\tau^2\psi\rangle$. Here, $\psi=(\psi_u,\psi_d)^T$ is a quark spinor in two-color and two-flavor system, $C=i\gamma^2\gamma^0$ is the charge-conjugation operator, and $\sigma^2$ and $\tau^2$ are the anti-symmetric Pauli matrices for color and flavor spaces, respectively. 

In the present study, in order to incorporate effects from the diquark condensate but investigate the CSE in the most transparent way, we start with the following concise effective model:
\begin{eqnarray}
{\cal L}_{\rm eff} = \bar{\psi}(i\Slash{D}+\mu\gamma^0-m)\psi+\psi^T{\bm \Delta}\psi\ . \label{LStart}
\end{eqnarray}
In this Lagrangian a covariant derivative $D_\mu\psi = (\partial_\mu-ieQA_\mu)\psi$ with $Q={\rm diag}(+\frac{2}{3},-\frac{1}{3})$ the charge matrix has been included to describe interactions with the external magnetic field. $m$ is a quark mass and ${\bm \Delta} =  \gamma_5\sigma^2\tau^2\Delta$ is responsible for the diquark condensate. For convenience we rewrite Eq.~(\ref{LStart}) into the Nambu-Gorkov basis as
\begin{eqnarray}
{\cal L}_{\rm eff} &=& \bar{\Psi}\left(
\begin{array}{cc}
i\Slash{\partial}+\mu\gamma_0-m & \bar{\bf \Delta} \\
{\bf \Delta} & i\Slash{\partial}-\mu\gamma_0-m \\
\end{array}
\right) \Psi \nonumber\\
&&  + e\bar{\Psi}\left(
\begin{array}{cc}
Q\Slash{A} & 0 \\
0 & -Q\Slash{A} \\
\end{array}
\right) \Psi\ ,  \label{LEffNG}
\end{eqnarray}
where
\begin{eqnarray}
\Psi \equiv \frac{1}{\sqrt{2}}\left(
\begin{array}{c}
\psi \\
\psi_c \\
\end{array}
\right)\ , \ \bar{\Psi} \equiv \frac{1}{\sqrt{2}}\left(
\begin{array}{cc}
\bar{\psi}\,, & \bar{\psi}_c
\end{array}
\right)
\end{eqnarray}
are the Nambu-Gor'kov spinors ($\psi_c =C\bar{\psi}^T$). From the first line of Eq.~(\ref{LEffNG}), the inverse of the fermion propagator $S^{-1}(p_0,{\bm p})$ is read as
\begin{eqnarray}
iS^{-1}(p_0,{\bm p})  = \left(
\begin{array}{cc}
\Slash{p}+\mu\gamma^0-m & \bar{\bf \Delta} \\
{\bf \Delta} &\Slash{p}-\mu\gamma^0-m \\
\end{array}
\right)  \ ,
\end{eqnarray}
with $\bar{\bm \Delta} \equiv \gamma^0{\bm \Delta}^\dagger\gamma^0$. By taking the inverse of $S^{-1}(p_0,{\bm p})$, the fermion propagator is obtained as
\begin{eqnarray}
S(p_0,{\bm p}) = \left(
\begin{array}{cc}
S_{11}(p_0,{\bm p}) & \sigma^2\tau^2S_{12}(p_0,{\bm p}) \\
\sigma^2\tau^2S_{21}(p_0,{\bm p}) & S_{22}(p_0,{\bm p}) \\
\end{array}
\right)\ , \label{2SCPropagatorM}
\end{eqnarray}
where each component has the Dirac structure as
\begin{eqnarray}
S_{11}(p_0,{\bm p}) &=& \sum_{\xi={\rm p}, {\rm a}}S^\xi_{11}(p_0,{\bm p})\Lambda_\xi  \gamma^0\ , \nonumber\\
S_{12}(p_0,{\bm p}) &=& \sum_{\xi={\rm p}, {\rm a}}S^\xi_{12}(p_0,{\bm p})\Lambda_\xi  \gamma_5\ , \nonumber\\
S_{21}(p_0,{\bm p}) &=& \sum_{\xi={\rm p}, {\rm a}}S^\xi_{21}(p_0,{\bm p})\Lambda^C_\xi  \gamma_5\ , \nonumber\\
S_{22}(p_0,{\bm p}) &=& \sum_{\xi={\rm p}, {\rm a}}S^\xi_{22}(p_0,{\bm p})\Lambda^C_\xi  \gamma^0\ ,  \label{SDec}
\end{eqnarray}
with
\begin{eqnarray}
S^{\xi}_{11} (p_0,{\bm p})
	&=& i\left(\frac{|u_{\xi}({\bm p})|^2 \,}{\, p_0 - \eta_\xi \epsilon_{\xi}({\bm p}) \,} 
		+ \frac{\, |v_{\xi}({\bm p})|^2 \,}{\, p_0 + \eta_\xi \epsilon_{\xi}({\bm p}) \,} \right)\ ,\nonumber\\
S^{\xi}_{12}(p_0,{\bm p}) 
	&=&-i\left( \frac{\, u_{\xi}^*({\bm p}) v_{\xi}^*({\bm p}) \,}{\, p_0-\epsilon_{\xi}({\bm p}) \,}  
		-  \frac{\, u_{\xi}^*({\bm p}) v_{\xi}^*({\bm p}) \,}{\, p_0+ \epsilon_{\xi}({\bm p}) \,} \right) \ ,\nonumber\\
S^{ {\xi}}_{21}(p_0,{\bm p}) 
	&=& i\left(\frac{\, u_{\xi}({\bm p}) v_{\xi}({\bm p}) \,}{\, p_0-\epsilon_{\xi}({\bm p}) \,} 
		- \frac{\, u_{\xi}({\bm p}) v_{\xi}({\bm p}) \,}{\, p_0 + \epsilon_{\xi}({\bm p}) \,} \right)\ , \nonumber\\
S^{ \xi }_{22} (p_0,{\bm p}) 
	&=& i\left( \frac{\, |v_{\xi} ({\bm p})|^2 \,}{\, p_0- \eta_\xi \epsilon_{\xi} ({\bm p}) \,} 
		+ \frac{\, |u_{\xi} ({\bm p})|^2 \,}{\, p_0 + \eta_\xi \epsilon_{\xi} ({\bm p}) \,} \right)\,.   
\label{SEach}
\end{eqnarray}
Here we introduced $\eta_{\rm p} = +1$ and $\eta_{\rm a} = -1$.
In these expressions, we have defined the positive-energy and negative-energy projection operators $\Lambda_{\rm p}$ and $\Lambda_{\rm a}$ by
\begin{eqnarray}
\Lambda_{\xi} &=&\gamma_0 \frac{\, E_{\bm p}\gamma_0+ \eta_\xi \big( m+{\bm \gamma}\cdot{\bm p}\big) \,}{2E_{\bm p}} \,.
\end{eqnarray}
%
%
%
with $E_{\bm p} = \sqrt{|{\bm p}|^2+m^2}$, and $\Lambda_{{\rm p}({\rm a})}^C = \Lambda_{{\rm a}({\rm p})}$, and
\begin{eqnarray}
\epsilon_{\xi}({\bm p}) &=& \sqrt{(E_{\bm p}- \eta_\xi \mu)^2+|\Delta|^2} \,, 
\end{eqnarray}
are the dispersion relations for quasiparticles. The factors $u_{\rm p}({\bm p}) $, $v_{\rm p}({\bm p}) $,  $u_{\rm a}({\bm p}) $, and $v_{\rm a}({\bm p})$ satisfy relations
\begin{eqnarray}
|u_{\xi}({\bm p}) |^2 &=& \frac{1}{\, 2 \,} \left(1+ \frac{\, E_{\bm p}- \eta_\xi \mu \,}{\, \epsilon_{\xi}({\bm p}) \,} \right)\ ,  \nonumber\\
|v_{\xi}({\bm p}) |^2 &=& \frac{1}{\, 2 \,} \left(1 - \frac{\, E_{\bm p}- \eta_\xi \mu \,}{\, \epsilon_{\xi}({\bm p}) \,} \right)\ , 
\end{eqnarray}
and
\begin{eqnarray}
\!\!\!\!
|u_{\xi}({\bm p}) |^2+|v_{\xi}({\bm p}) |^2 = 1\,,~~
 u_{\xi}({\bm p})  v_{\xi}({\bm p})  = \frac{\Delta}{2\epsilon_{\xi}({\bm p}) } \,.
\end{eqnarray}
By using the propagator of the quasiparticles in Eq.~(\ref{2SCPropagatorM}) together with Eqs.~(\ref{SDec}) and~(\ref{SEach}), and with the help of linear response theory, the CSE in the diquark condensed phase under a weak magnetic field can be evaluated.

\section{CS conductivity}
\label{sec:CSE_conductivity}

In the present work we investigate the CSE in the diquark condensed phase within the linear response theory. Our calculations are similar to that for the chiral magnetic conductivity \cite{Kharzeev:2009pj}, but our calculations are more complicated due to the presence of diquark condensates. 
The axial current is defined by $j_5^i = \bar{\psi}\gamma^i\gamma_5\psi$ ($i=1,2,3$). 
In the linear response theory we measure the difference between the current with and without external electromagnetic vector potential $A$,
\begin{eqnarray}
\langle j_5^i \rangle \equiv \langle j_5^i \rangle_{A} - \langle j_5^i \rangle_{A=0} \,.
\end{eqnarray}
We note that the $U(1)_A$ axial current couples to the gluon topological density as $\partial^\mu j_\mu^A \sim \tilde{G}_{\mu \nu} G^{\mu \nu}$.
Below we neglect the impact of $A$ on the color gauge dynamics, i.e., $\langle \tilde{G}_{\mu \nu} G^{\mu \nu} \rangle_{A} = \langle \tilde{G}_{\mu \nu} G^{\mu \nu} \rangle_{A=0}$. 
Then we focus on the anomalous contributions coupled to the $\mu$ and $A$.

In the following calculation, we will make use of the imaginary time formalism to incorporate finite temperature effects. In momentum space the axial current induced by the external electromagnetic field is evaluated as the retarded correlators between the axial vector and vector currents, 
($\int_{\bm p } \equiv \int d^3 p/(2\pi)^3$)
\beq
\langle j_5^i(i\bar{\omega}_n,{\bm q})\rangle 
&\equiv& \int_0^\beta d\tau \int d^3x\, \langle \bar{\psi}(x)\gamma^i\gamma_5\psi(x)\rangle {\rm e}^{-i\bar{\omega}_n \tau+i{\bm q}\cdot{\bm x}} \nonumber\\
 &=& \frac{\, e \,}{2} \int_{\bm p} T \sum_m 
  {\rm Tr}\Big[\Gamma^i\Gamma_5  {S}(i\omega_m,{\bm p}) \nonumber\\
 &\times&\big(\Gamma^{{\rm bare}}_{\bar{\psi}A\psi}\big)^\mu{S}(i\omega_m',{\bm p}')\Big]{A}_\mu(i\bar{\omega}_n,{\bm q})\, ,  \label{CSE1}
\eeq
with ${\bm p}'={\bm p}+{\bm q}$, $i\omega_m' = i\omega_m+i\bar{\omega}_n$ and $N_c=2$, where we have defined the vertices in the Nambu-Gor'kov space as
\begin{eqnarray}
\Gamma^i\Gamma_5 =  \left(
\begin{array}{cc}
\gamma^i \gamma_5 & 0 \\
0 & \gamma^i \gamma_5 \\
\end{array}
\right)\ , \big(\Gamma^{{\rm bare}}_{\bar{\psi}A\psi}\big)^\mu = \left(
\begin{array}{cc}
Q\gamma^\mu & 0 \\
0 & -Q\gamma^\mu \\
\end{array}
\right)\ . \nonumber\\
\end{eqnarray}
Here $\omega_m = (2m+1)\pi T$ and $\bar{\omega}_n=2n\pi T$ ($n$, $m$ are integers) are the Matsubara frequencies for fermions and bosons, respectively. The symbol ``Tr'' stands for the trace with respect to the color, flavor, spinor, and Nambu-Gor'kov spaces. (We note that, in the presence of symmetry breaking, the use of the bare vertices violates the conservation law and one must improve the vertices to recover it. We will come back to this point later, just mentioning that such improvements will not change the main conclusion of the simplest one loop results.)

We will leave only spatial components of the external gauge field $A_\mu(i\bar{\omega}_n,{\bm q})$ because the magnetic field is solely generated by them. The real time axial current $\langle j_5^i(q_0,{\bm q})\rangle $ is given by the analytic continuation as
\begin{eqnarray}
\langle j_5^i(q_0,{\bm q})\rangle = \langle j_5^i(i\bar{\omega}_n,{\bm q})\rangle\Big|_{i\bar{\omega}_n\to q_0+i\eta}\ , \label{AnCont}
\end{eqnarray}
with $\eta$ an infinitesimal positive number.

Performing the trace with respect to the color, spinor, and Nambu-Gorkov spaces in Eq.~(\ref{CSE1}), the axial current is reduced to the form
\begin{eqnarray}
\!\!\!\!\!\!\!\!\!\!\!\!
\langle j_5^i(i\bar{\omega}_n,{\bm q})\rangle 
&=& -i e \,{\rm tr}[Q] \epsilon^{ijk} A^j(i\bar{\omega}_n,{\bm q}) \sigma_{\rm CSE}^k(i\bar{\omega}_n,{\bm q})
\nonumber\\
&=& -e \,{\rm tr}[Q] B^i   \sigma_{\rm CSE} (i\bar{\omega}_n,{\bm q})
\end{eqnarray}
where we used that $\sigma^k_{\rm CSE}(i\bar{\omega}_n,{\bm q}) = q^k \sigma_{\rm CSE}(i\bar{\omega}_n,{\bm q}) $ because of the Lorentz structure, and $B^i = \epsilon^{imn} (-i q^m) A^n $.
Hence we can extract the conductivity $\sigma_{\rm CSE}(i\bar{\omega}_n,{\bm q})$ as 
\begin{eqnarray}
\sigma_{\rm CSE}(i\bar{\omega}_n,{\bm q}) = \frac{\, q^k \,}{\, |{\bm q}|^2 \,} \, \sigma^k_{\rm CSE}(i\bar{\omega}_n,{\bm q}) \,.
\end{eqnarray}
The function $\sigma^k_{\rm CSE}$ is computed using propagators decomposed into the particle and antiparticle pieces with normal and anomalous components, as in Eq.(\ref{SDec}).
Accordingly the function $\sigma_{\rm CSE}$ is also decomposed as
\begin{eqnarray}
\sigma_{\rm CSE}(i\bar{\omega}_n,{\bm q}) = \sum_{\xi,\xi'={\rm p},{\rm a}}  \sigma^{\xi \xi'}_{\rm CSE}(i\bar{\omega}_n,{\bm q}) \,.
\end{eqnarray}
The set $(\xi, \xi') = ({\rm p}, {\rm p})$ and $({\rm a}, {\rm a})$ correspond to the particle-hole and antiparticle-antihole contributions, while $({\rm p}, {\rm a})$ or $({\rm a}, {\rm p})$ are from the particle-antiparticle contributions.
Explicitly,
\begin{eqnarray}
\!\!\!\!\!\!\!\!
 \sigma^{\xi \xi'}_{\rm CSE}(i\bar{\omega}_n,{\bm q}) 
= \frac{\, N_c q^k \,}{ |{\bm q}|^2} \int_{\bm p} {\cal F}^{\xi \xi'} ( \bar{\omega}_n, {\bm q}; {\bm p} ) {\cal K}^k_{\xi \xi'} ({\bm q}; {\bm p}) \,,
\end{eqnarray}
%
%
with the kinematic factor
\begin{eqnarray}
{\cal K}^k_{\xi \xi'} ({\bm q}; {\bm p}) = - \eta_{\xi}\frac{ p^k }{E_{\bm p}} + \eta_{\xi'}\frac{ (p')^{k}}{E_{{\bm p}'}} \,,
\end{eqnarray}
which are independent of the diquark condensates,
and the propagator part
\begin{widetext}
\begin{eqnarray}
{\cal F}^{\xi \xi'} ( \bar{\omega}_n, {\bm q}; {\bm p} )
= T\sum_m \Big[
	S_{11}^\xi(i\omega_m,{\bm p})S_{11}^{\xi'}(i\omega'_m,{\bm p}')
	- S_{12}^\xi(i\omega_m,{\bm p}) S_{21}^{\xi'}(i\omega'_m,{\bm p}')\nonumber\\
-  S_{21}^{\xi}(i\omega_m,{\bm p})S_{12}^{\xi'}(i\omega'_m,{\bm p}')
	+S_{22}^{\xi}(i\omega_m,{\bm p})S_{22}^{\xi'}(i\omega'_m,{\bm p}')
\Big] \,,
\end{eqnarray}
\end{widetext}
which carries the effects from the diquark condensates.
In ${\cal F}^{\xi \xi'}$ we can carry out the Matsubara summation using an identity
\begin{eqnarray}
T\sum_m \frac{1}{(i\omega_m-\epsilon_1)(i\omega_m'-\epsilon_2)} = \frac{\tilde{f}(\epsilon_1)-\tilde{f}(\epsilon_2)}{i\bar{\omega}_n-\epsilon_2+\epsilon_1}\ , \label{MatsubaraSum}
\end{eqnarray}
where $\tilde{f}(\epsilon)  = 1/({\rm e}^{\beta \epsilon}+1)$ is the Fermi-Dirac distribution function. 
Using this identity, the function ${\cal F}^{\xi \xi'}$ can be written as
\begin{eqnarray}
&&{\cal F}^{\xi \xi'} ( \bar{\omega}_n, {\bm q}; {\bm p} )
\nonumber \\
&& = {\cal C}_1^{\xi \xi'} ({\bm q};{\bm p}) \,  {\cal G}_1^{\xi \xi'} (q;p) + {\cal C}_2^{\xi \xi'} ({\bm q};{\bm p}) \, {\cal G}_2^{\xi \xi'} (q;p) \,,
\end{eqnarray}
where the second term is nonzero only at finite temperature. The factors ${\cal C}^{\xi \xi'}$ and $\tilde{ {\cal C} }^{\xi \xi'}$ are the coherence factors which carry the information about the wavefunctions $u_\xi$ and $v_\xi$, 
%
\beq
{\cal C}_{1,2}^{ {\rm p} {\rm p} } ({\bm q} ; {\bm p}) 
 & =  \frac{1}{2}\left[1 \mp \frac{\, (E_{\bm p} - \mu)(E_{{\bm p}'} - \mu) +  |\Delta|^2 \,}{ \epsilon_{\rm p} ({\bm p})\epsilon_{\rm p} ({\bm p}')} \right] \, ,
\nonumber \\
{\cal C}_{1,2}^{ {\rm a} {\rm a} } ({\bm q} ; {\bm p}) 
 & =  \frac{1}{2}\left[1 \mp \frac{\, (E_{\bm p} + \mu)(E_{{\bm p}'} + \mu) +  |\Delta|^2 \,}{ \epsilon_{\rm a} ({\bm p})\epsilon_{\rm a} ({\bm p}')} \right] \, ,
\nonumber \\
{\cal C}_{1,2}^{ {\rm p} {\rm a} } ({\bm q} ; {\bm p}) 
 & =  \frac{1}{2}\left[1 \pm \frac{\, (E_{\bm p} - \mu)(E_{{\bm p}'} + \mu) -  |\Delta|^2 \,}{ \epsilon_{\rm p} ({\bm p})\epsilon_{\rm a} ({\bm p}')} \right] \, ,
 \nonumber \\
{\cal C}_{1,2}^{ {\rm a} {\rm p} } ({\bm q} ; {\bm p}) 
 & =  \frac{1}{2}\left[1 \pm \frac{\, (E_{\bm p} + \mu)(E_{{\bm p}'} - \mu) -  |\Delta|^2 \,}{ \epsilon_{\rm a} ({\bm p})\epsilon_{\rm p} ({\bm p}')} \right] \, .
\label{coherence_factor}
\eeq
%
%
The upper (lower) sign in the right-hand side (RHS) corresponds to the subscript $1$ ($2$) in the left-hand side (LHS). The factors ${\cal G}_{1,2}^{\xi \xi'}$ are the propagator factors which carry the information about the excitation spectra,
\beq
&& {\cal G}_1^{\xi \xi'} (q;p) 
=  \frac{\, 1  \,}{2} \big[\, 1 - \tilde{f} \big(\epsilon_\xi ({\bm p}) \big)  - \tilde{f} \big(\epsilon_{\xi'} ({\bm p}' ) \big) \, \big] 
\nonumber \\
&& 
\times \bigg( \frac{1}{\, i\bar{\omega}_n + \epsilon_{\xi}({\bm p}) + \epsilon_{\xi'} ({\bm p}') \,} 
+ \frac{1}{\, - i \bar{\omega}_n + \epsilon_{\xi}({\bm p}) + \epsilon_{\xi'} ({\bm p}') \,} \bigg) \,,
\nonumber \\
\eeq
and 
\begin{eqnarray}
&& {\cal G}_2^{\xi \xi'} (q;p)
= - \frac{\, 1 \,}{2} \big[\, \tilde{f}\big(\epsilon_{\xi} ({\bm p})\big) - \tilde{f}\big(\epsilon_{\xi'}({\bm p}')\big) \,\big]
\nonumber \\
&& \times \bigg( \frac{1}{\, i\bar{\omega}_n +\epsilon_{\xi}({\bm p}) -\epsilon_{\xi'} ({\bm p}') \,} 
+ \frac{1}{\, -i\bar{\omega}_n +\epsilon_{\xi}({\bm p}) -\epsilon_{\xi'} ({\bm p}') \,} \bigg) \,.
\nonumber \\
\end{eqnarray}
Below we examine how the contributions in different sectors affect the CSE conductivity.

\subsection{Particle-hole}

The coherence factors and propagators behave very differently for the normal and condensed phases. We consider the low temperature case $T \ll \Delta$ where quantum effects are important.
The contributions which are most sensitive to the condensates are particle-hole contributions near the edge of the Fermi sea with $| {\bm p}| \simeq | {\bm p}' | \simeq p_F$.
Below we discuss the particle-hole contributions for the $|{\bm q}| \ll \Delta$ and $|{\bm q}| \gg \Delta$ cases. The latter can be seen as the results in the normal phase.

\subsubsection{ $ | {\bf q} | \ll \Delta$ }

Depending on the size of ${\bm q}$, the coherence factors appear constructively or destructively.
To see this, first we set ${\bm q} = 0$ as a special case of $|{\bm q}| \ll \Delta$. Then we find
\beq
{\cal C}_1^{ {\rm p} {\rm p} } ( {\bm q}=0; {\bm p} ) = 0 \,,~~~~~ {\cal C}_2^{ {\rm p} {\rm p} } ( {\bm q}=0; {\bm p} ) = 1 \,.
\eeq
For ${\cal C}_1^{ {\rm p} {\rm p} } $, the correction starts with $O({\bm q})$. 
With this ${\cal F}^{ {\rm p} {\rm p} } \sim {\cal G}_2^{ {\rm p} {\rm p} } $ for small ${\bm q}$.
For $ {\cal G}_2^{ {\rm p} {\rm p} }$, we focus on $| {\bm p}| \simeq | {\bm p}' | \simeq p_F$, where $\epsilon_{\rm p} (p_F) = \Delta$ and $\tilde{f} \sim {\rm e}^{-\Delta/T} \ll 1$. 
We expand
\beq
\!\!\!\!\!\!\!\!
 \tilde{f}\big(\epsilon_{\rm p} ({\bm p})\big) - \tilde{f}\big(\epsilon_{\rm p}({\bm p}')\big)
 \simeq
 -  \frac{\, \partial \tilde{f} (x) \,}{\, \partial x \,} \bigg|_{x=\epsilon_{\rm p} }   \frac{\, \partial \epsilon_{\rm p} \,}{\,\partial {\bm p} \,}  \cdot {\bm q} \,,
\eeq
and get
\beq
&&\!\!\!\!\!\!\!\!
 {\cal G}_2^{ {\rm p} {\rm p} } 
\sim   
 \frac{\, \partial \tilde{f} (x) \,}{\, \partial x \,} \bigg|_{x=\epsilon_{\rm p} }  
   \frac{\, \big( \frac{\, \partial \epsilon_{\rm p} \,}{\,\partial {\bm p} \,}  \cdot {\bm q} \big)^2 \,}
	{\, \bar{\omega}_n^2 + \big( \frac{\, \partial \epsilon_{\rm p} \,}{\,\partial {\bm p} \,}  \cdot {\bm q} \big)^2 \,} 
~\sim~ \frac{1}{T} {\rm e}^{-\Delta/T} \,.
	\label{eq:F_condensate}
\eeq
For $T\rightarrow 0$, $\partial \tilde{f}/\partial x \rightarrow 0$ at $x >0$, and $ {\cal G}_2^{ {\rm p} {\rm p} } $ is vanishing for $ \Delta \neq 0$ as ${\rm e}^{-\Delta/T}/T\rightarrow 0$.

Finally we look at the kinematic factor couples to $q^k$,
\begin{eqnarray}
&&q^k {\cal K}^k_{ {\rm p} {\rm p} } ({\bm q}; {\bm p}) 
= - \frac{ {\bm q} \cdot {\bm p} }{E_{\bm p}} + \frac{ {\bm q} \cdot ( {\bm p}+ {\bm q}) }{E_{ {\bm p} +{\bm q} }}
\nonumber \\
&& 
 \simeq \frac{\, {\bm q}^2 \,}{\, E_{\bm p} \,} - \frac{\, ( {\bm p} \cdot {\bm q} )^2 \,}{\, E_{\bm p}^3 \,} 
~ \sim~ \frac{\, {\bm q}^2 \,}{\, E_{\bm p} \,} \bigg( 1 - \frac{1}{3} \frac{\,  |{\bm p}|^2 \,} {\, E_{\bm p}^2 \,} \bigg)
 \,,
\end{eqnarray}
where in the last expression we averaged $  ( {\bm p} \cdot {\bm q} )^2 \rightarrow {\bm p}^2 {\bm q}^2/3$ in the integral over ${\bm p}$. Then the CSE conductivity from the particle-hole can be written as
\beq
\!\!\!\!\!\!\!\!
 \sigma^{ {\rm p} {\rm p} }_{\rm CSE}(i\bar{\omega}_n, {\bm q}) 
~\sim~ N_c \int_{\bm p} {\cal G}_2^{ {\rm p} {\rm p} } 
	\frac{\, 1 \,}{\, E_{\bm p} \,} \bigg( 1 - \frac{1}{3} \frac{\,  |{\bm p}|^2 \,} {\, E_{\bm p}^2 \,} \bigg) \,.
\label{eq:CSE_pp_qsmall}
\eeq
for small ${\bm q}$. Note that the thermal factor makes the expression Eq.(\ref{eq:CSE_pp_qsmall}) UV finite.

At zero temperature the particle-hole contributions vanish by the presence of the diquark gap as ${\cal G}_2^{ {\rm p} {\rm p} } \sim {\rm e}^{-\Delta/T} /T\rightarrow 0$. 
This suppression of particle-hole is what happens for the Meissner effect in a superconductor. 
In contrast, in a normal conductor the particle-hole induces the paramagnetic effects which cancel with the diamagnetic contributions from particle-antiparticle contributions.
The absence of particle-hole contributions or paramagnetic effects, as found in our calculations for the diquark condensed phase, makes the material diamagnetic. The particle-hole pairs contribute only as thermal excitations.

\subsubsection{ $ | {\bf q} | \gg \Delta$ or normal phase }

For $|{\bm q}| \gg \Delta$, the above expression is qualitatively modified. Here it is important to note that $ (E_{\bm p} - \mu)(E_{{\bm p}'} - \mu) < 0 $ for the particle-hole contributions; one of the excitations has $E>\mu$ and the other $E <\mu$. Then, neglecting $\Delta$, one finds
\beq
 \frac{\, (E_{\bm p} - \mu)(E_{{\bm p}'} - \mu) + |\Delta|^2 \,}{\epsilon_{\rm p} ({\bm p})\epsilon_{\rm p} ({\bm p}')} ~\simeq~ - 1 \,,
\eeq
so that
\beq
{\cal C}_1^{ {\rm p} {\rm p} } ( {\bm q}; {\bm p} ) ~\simeq~ 1 \,,~~~~~ {\cal C}_2^{ {\rm p} {\rm p} } ( {\bm q}; {\bm p} ) ~\simeq~  0 \,.
\eeq
With this ${\cal F}^{ {\rm p} {\rm p} } \sim {\cal G}_1^{ {\rm p} {\rm p} } $ for $ |{\bm q}| \gg \Delta$. This function is evaluated by focusing on the domain $| {\bm p}| \simeq | {\bm p}' | \simeq p_F$. For $E_{\bm p} - \mu \gg \Delta$ and  $\mu-E_{{\bm p}'} \gg \Delta$, we use
\beq
1 - \tilde{f} \big(\epsilon_{\rm p} ({\bm p}) \big)  - \tilde{f} \big(\epsilon_{\rm p} ({\bm p}' ) \big)
&~\sim~&
 	\tilde{f} \big(E_{ {\bm p}' } - \mu\big) - \tilde{f} \big(E_{ {\bm p} } - \mu\big) 
 \nonumber \\
 &~\sim~&
	  \frac{\, \partial \tilde{f} (x) \,}{\, \partial x \,} \bigg|_{x= E_{\bm p}-\mu }    \frac{\, \partial E_{\bm p} \,}{\,\partial {\bm p} \,}  \cdot {\bm q} \,,
  	\nonumber \\
&&\hspace{-4cm} 
\epsilon_{\xi}({\bm p}) + \epsilon_{\xi'} ({\bm p}') 
~\sim~ E_{\bm p} - E_{{\bm p}'} 
~\simeq~ - \frac{\, \partial E_{\bm p} \,}{\,\partial {\bm p} \,}  \cdot {\bm q} \,,
\eeq
to get
\beq
\!\!\!\!\!\!\!\!
 {\cal G}_1^{ {\rm p} {\rm p} } 
\, \sim \,   
- \frac{\, \partial \tilde{f}(x) \,}{\, \partial x \,} \bigg|_{x= E_{\bm p}-\mu } 
  \times \frac{\, \big( \frac{\, \partial E_{\rm p} \,}{\,\partial {\bm p} \,}  \cdot {\bm q} \big)^2 \,}
	{\, \bar{\omega}_n^2 + \big( \frac{\, \partial E_{\rm p} \,}{\,\partial {\bm p} \,}  \cdot {\bm q} \big)^2 \,} \,.
\label{eq:F_normal}
\eeq
For $T\rightarrow 0$, the contribution entirely comes from $|{\bm p}| = p_F$ where $\partial \tilde{f}/ \partial x $ yields $ - \delta( |{\bm p}|-p_F) $, as in the usual Debye screening calculations. 
Including the kinematic factor as before, we find
\beq
\!\!\!\!\!\!\!\!
 \sigma^{ {\rm p} {\rm p} }_{\rm CSE}(i\bar{\omega}_n, {\bm q}) 
~\sim~ N_c \int_{\bm p} {\cal G}_1^{ {\rm p} {\rm p} }
	\frac{\, 1 \,}{\, E_{\bm p} \,} \bigg( 1 - \frac{1}{3} \frac{\,  |{\bm p}|^2 \,} {\, E_{\bm p}^2 \,} \bigg)  \,,
\label{eq:CSE_pp_qlarge}
\eeq
for $|{\bm q}| \gg \Delta$. The expression Eq.(\ref{eq:CSE_pp_qlarge}) is UV finite and the integrand is dominated by the contributions from $|{\bm p}|\simeq p_F$. 

As mentioned before, the result for $|{\bm q}| \gg \Delta$ can be taken as a result for the normal phase with $\Delta = 0$. In the normal phase the particle-hole contributions are large at $T=0$ as they are gapless, in contrast to fermions in the diquark condensed phase.
In the massless limit, we find, for the static limit,
\beq
 \sigma^{ {\rm p} {\rm p} ({\rm normal}) }_{\rm CSE;\, {\it m}=0 } \big|^{\omega =0}_{ {\bm q} \rightarrow 0 } 
	~ \rightarrow ~ \frac{\, N_c \mu \,}{\, 3\pi^2 \,} \,.
\eeq
and for the dynamic limit
\beq
 \sigma^{ {\rm p} {\rm p} ({\rm normal}) }_{ {\rm CSE};\, {\it m}=0 }\big|^{\omega \rightarrow 0}_{ {\bm q} = 0 } 
	~ \rightarrow ~ 0 \,.
\eeq
%

\subsection{Particle-antiparticle}


Next we discuss the particle-antiparticle contributions. These contributions are not regulated by thermal factors and are potentially UV divergent. Our main concern is the UV finiteness and for this purpose it is sufficient to study $T \sim 0$.

The coherence factors in the particle-antiparticle contributions are not dominated by the contributions from $|{\bm p} | \sim |{\bm p}' | \simeq p_F$, but come from large phase space with momenta much larger than $\Delta$. For this reason we set $\Delta \rightarrow 0$ and focus on the leading order contributions. Then for $E_{\bm p} -\mu>0$,
\beq
 \frac{\, (E_{\bm p} - \mu)(E_{{\bm p}'} + \mu) - |\Delta|^2 \,}{\epsilon_{\rm p} ({\bm p})\epsilon_{\rm a} ({\bm p}')} ~\rightarrow~ 1 \,,
\eeq
Accordingly,
\beq
{\cal C}_1^{ {\rm p} {\rm a} } ( {\bm q}; {\bm p} ) ~\rightarrow~ 1 \,,~~~~~ {\cal C}_2^{ {\rm p} {\rm a} } ( {\bm q}; {\bm p} ) ~\rightarrow~  0 \,,
\eeq
and we find
\beq
\!\!\!\!\!\!\!\!\!
{\cal F}^{ {\rm p} {\rm a} } 
\rightarrow \,
{\cal G}^{ {\rm p} {\rm a} }_1
= \big[\, 1 - \tilde{f} \big(E_{\bm p} - \mu \big) - \tilde{f} \big(E_{{\bm p}'} + \mu \big) \, \big] G_{ {\rm p} {\rm a} }  ,
\eeq
where
\beq
G_{ {\rm p} {\rm a} }
= \frac{\, E_{\bm p} + E_{{\bm p}'} \,}{\, \bar{\omega}_n^2 + \big( E_{\bm p} + E_{{\bm p}'} \big)^2 \,} 
 \,.
\eeq
Likewise
\beq
\!\!\!\! 
{\cal F}^{ {\rm a} {\rm p} } 
 \rightarrow 
{\cal G}^{ {\rm a} {\rm p} }_1
= \big[\, 1 - \tilde{f} \big(E_{\bm p} + \mu \big) - \tilde{f} \big(E_{{\bm p}'} - \mu \big) \, \big] G_{ {\rm p} {\rm a} }.
\eeq
Meanwhile the kinematic factor is
\begin{eqnarray}
{\cal K}^k_{ {\rm p} {\rm a} } = - {\cal K}^k_{ {\rm a} {\rm p} } = - \bigg( \frac{ p^k }{E_{\bm p}} + \frac{ (p')^{k}}{E_{{\bm p}'}} \bigg) 
 \sim 
- 2  \frac{\, p^k \,}{\, E_{\bm p} \,} \,,
\end{eqnarray}
for small ${\bm q}$.
While each of (pa) and (ap) contributions is UV divergent, the sum cancels the UV divergence.
\beq
&& {\cal F}^{ {\rm p} {\rm a} }  {\cal K}^k_{ {\rm p} {\rm a} } +  {\cal F}^{ {\rm a} {\rm p} } {\cal K}^k_{ {\rm a} {\rm p} } 
\, \sim \, 
 {\cal K}^k_{ {\rm p} {\rm a} }  \big(  {\cal F}^{ {\rm p} {\rm a} } -  {\cal F}^{ {\rm a} {\rm p} } \big)
\nonumber \\
&&
\, \sim \, 
 - 2 \frac{\, p^k \,}{\, E_{\bm p} \,} \, G_{ {\rm p} {\rm a} }
	\big[\, \tilde{f} \big(E_{ {\bm p}'} - \mu \big) - \tilde{f} \big(E_{\bm p} - \mu \big)  \,\big]  - (\mu \leftrightarrow-\mu)
\nonumber \\
&& \sim  
- 2 \frac{\, p^k  \,}{\, E_{\bm p} \,}  \frac{\, ( {\bm p}\cdot {\bm q} )  \,}{\, E_{\bm p} \,} \, G_{ {\rm p} {\rm a} }
\sum_{ \xi = {\rm p}, {\rm a} } \eta_\xi
	 \frac{\, \partial \tilde{f} (x) \,}{\, \partial x \,} \bigg|_{x= E_{\bm p} - \eta_\xi \mu }   
\nonumber \\
&& \sim 
 - \frac{\, 2 q^k \,}{3} \frac{\, {\bm p}^2 \,}{\, E_{\bm p}^2 \,} 	\, G_{ {\rm p} {\rm a} }
 \sum_{ \xi = {\rm p}, {\rm a} } \eta_\xi
	 \frac{\, \partial \tilde{f} (x) \,}{\, \partial x \,} \bigg|_{x= E_{\bm p}- \eta_\xi \mu }  \,,
\eeq
where we took the angular average for ${\bm p}$. The CS conductivity is
\beq
&&\!\!\!\! \sigma^{ {\rm p} {\rm a} + {\rm a} {\rm p} }_{\rm CSE}(i\bar{\omega}_n, {\bm q}) 
\nonumber \\
&& \!\!\!\!
\sim \,- N_c \int_{\bm p} 
	\frac{\, 2 \,}{3} \frac{\, {\bm p}^2 \,}{\, E_{\bm p}^2 \,} \, G_{ {\rm p} {\rm a} }	 
 	\sum_{ \xi = {\rm p}, {\rm a} } \eta_\xi
		 \frac{\, \partial \tilde{f} (x) \,}{\, \partial x \,} \bigg|_{x= E_{\bm p}- \eta_\xi \mu }  \,.
\label{eq:CSE_pa_qlarge}
\eeq
Because of $\partial \tilde{f}/ \partial x $, the integrand is dominated by states with $E_{\bm p} \pm \mu$.
In the massless limit, we find, for the static limit,
\beq
 \sigma^{ {\rm p} {\rm p} ({\rm normal}) }_{ {\rm CSE};\, {\it m}=0 } \big|^{\omega =0}_{ {\bm q} \rightarrow 0 } 
	~ \rightarrow ~ \frac{\, N_c \mu \,}{\, 6 \pi^2 \,} \,.
\eeq
and for the dynamic limit
\beq
 \sigma^{ {\rm p} {\rm p} ({\rm normal}) }_{ {\rm CSE};\, {\it m}=0} \big|^{\omega \rightarrow 0}_{ {\bm q} = 0 } 
	~ \rightarrow ~ \frac{\, N_c \mu \,}{\, 6 \pi^2 \,} \,.
\eeq
That is, the static and dynamic limits are the same, as the propagators do not have the sensitivity to the limiting order.

\subsection{Antiparticle-antihole}

The computations of antiparticle-antihole proceed in the very similar way as the particle-hole case. The contributions are suppressed by thermal factor as $\sim {\rm e}^{-(\mu+\Delta)/T}$.

\section{Numerical results}
\label{sec:Results}

In this section, we present the numerical results of the CS conductivity for plural density, quark mass, and temperature. We discuss a normalized CS conductivity,
\beq
R^{\xi \xi'} \equiv \sigma^{\xi \xi'}_{\rm CSE}/\sigma_0 \,,~~~~~ R = \sum_{\xi,\xi'={\rm p},{\rm a} } R^{\xi \xi'} \,,
\eeq
with a $\mu$ dependent normalization factor
\beq
\sigma_0 = \frac{\, N_c \mu \,}{\, 2\pi^2 \,} \,,
\eeq
where $\sigma_0 $ is the conductivity at massless, zero temperature, and static limits in the normal phase.

To understand the results in this section, it is useful to use the massless and zero temperature limit in the normal phase as a baseline. The static conductivity is
\beq
\!\!\!\!\!\!\!\!
\big( R^{{\rm p} {\rm p} } + R^{{\rm p} {\rm a} + {\rm a} {\rm p} } + R^{{\rm a} {\rm a} } \big)^{\omega=0}_{ {\bm q} \rightarrow 0 }
\, \rightarrow \,
 \frac{2}{\, 3 \,} + \frac{1}{\, 3 \,}  + 0 = 1 \,,
\eeq
and the dynamic conductivity is
\beq
\!\!\!\!\!\!\!\!
\big( R^{{\rm p} {\rm p} } + R^{{\rm p} {\rm a} + {\rm a} {\rm p} } + R^{{\rm a} {\rm a} } \big)^{\omega \rightarrow 0}_{ {\bm q} = 0 }
\, \rightarrow \,
0 + \frac{1}{\, 3 \,} + 0 = \frac{1}{\, 3 \,} \,.
\eeq
The difference comes from the response of particle-hole to the external fields. 

To delineate the CS conductivity we found it convenient to discuss the normal phase as a guideline. 
We discuss the CS conductivity including the mass effects, $\mu$ dependence, and temperature dependence. Finally we include the diquark gap. We focus on the $\omega = 0$ case; the impact of $\omega$ can be readily seen from the propagator factors.

In this section we present the CS conductivity pretending that quark matter exists from $\mu \ge m$. This picture is not realistic at $\mu \sim m$ for QC$_2$D where a hadronic phase should exist at $\mu < m_\pi/2$ and diquarks with the mass $m_\pi$ condense at $\mu\ge m_\pi/2$.  At low density there are more appropriate presentations based on the chiral Lagrangian \cite{Avdoshkin:2017cqp}, and we will not repeat those discussions.
Hence, only our results at $\mu \gg m$ can be taken at its face value. Here the chiral symmetry is supposed to get restored and $m$ will be regarded as the current quark mass. As for comparisons with the lattice simulation setup, we note that the pion mass used is about $700$ MeV, and we should regard the current quark mass as rather heavy. We take $m\sim 100$ MeV as a typical value.

\subsection{Normal phase}
\label{sec:normal_phase}

\begin{figure}[tbp]
\centering
\includegraphics*[scale=0.8]{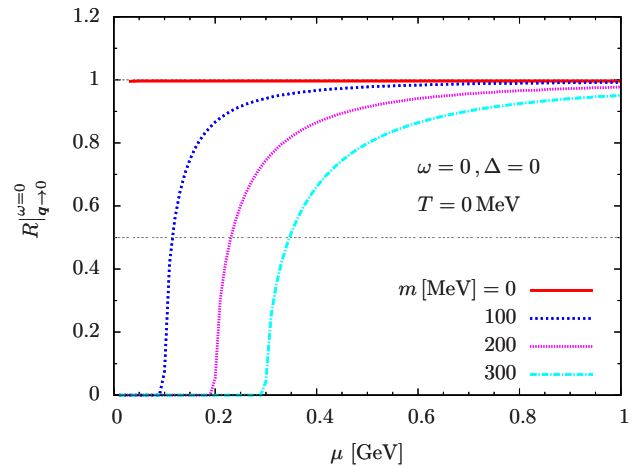}
\caption{The $\mu$ dependence of the CS conductivity for the quark masses $m =0, 100, 200,$ and 300 MeV, where $\omega, T, \Delta$ are set to zero. 
}
\label{fig:FIG_Mq_dep_del00_T1_mu_dep_k0}
\end{figure}

First we examine the current quark mass dependence.
A finite current quark mass breaks the chiral symmetry explicitly, blurring qualitative pictures for chiral conductivities whose descriptions are usually based on massless quarks.
Whether the mass effects enter as a quark mass or the mass of NG modes depend on the phase structure of the QC$_2$D, 
but we consider the former case as a guide for the domain of $\mu \gg m$.
Shown in Fig.\ref{fig:FIG_Mq_dep_del00_T1_mu_dep_k0} is the (normalized) static CS conductivity at zero temperature.
It sets the overall size. The cases with various quark masses $m$ are displayed. Clearly the mass characterizes the onset of quark density. We have checked the scaling
\beq
\sigma_{\rm CSE} \big|^{\omega=0}_{ {\bm q} \rightarrow 0 } 
= \frac{\, N_c \sqrt{\mu^2 - m^2 \,} \,}{\, 2\pi^2 \,} = \frac{\, N_c p_F \,}{\, 2\pi^2 \,} \,.
\eeq
where $p_F$ is the Fermi momentum in the normal phase, related to the baryon density as $p_F=(3\pi^2 n_B/N_f)^{1/3}$.
This expression is useful to understand that the CS conductivity is sensitive to the number density, rather than the chemical potential. Below we set $m=0$.

\begin{figure}[htbp]
\centering
\includegraphics*[scale=0.8]{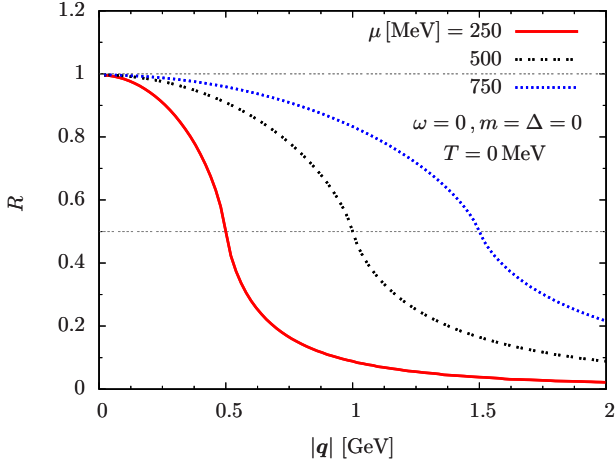}
\caption{ The $\mu$ dependence of the CS conductivity as a function of spatial momenta ${\bm q}$. As an eye-guide we show a $R=0.5$ line which is found at $|{\bm q}|=2\mu =2p_F$.
}
\label{fig:Mq0.0_del0.0_T0.01_mu_dep}
\end{figure}

Next we examine the $\mu$ dependence of the normalized CS conductivity $R$ at various external momenta $|{\bm q}|$, shown in Fig.\ref{fig:Mq0.0_del0.0_T0.01_mu_dep}. The $R$ in low momentum limit approaches $1$ but damps at larger momenta. Its size becomes half at $|{\bm q}| = 2\mu = 2p_F$, but after that the damping proceeds rather slowly.

\begin{figure}[htbp]
\centering
\includegraphics*[scale=0.8]{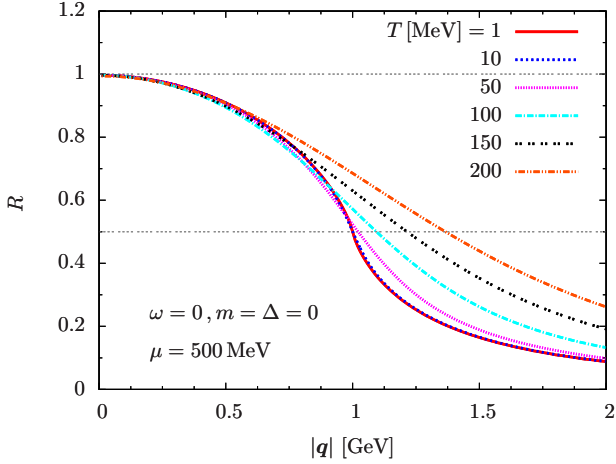}
\caption{ The $T$ dependence of the CS conductivity as a function of spatial momenta ${\bm q}$.  
}
\label{fig:Mq00_del00_mu05_T_dep}
\end{figure}

Shown in Fig.\ref{fig:Mq00_del00_mu05_T_dep} is the temperature dependence of $R$. We fixed $\mu=500$ MeV. 
The $R$ at ${\bm q}\rightarrow 0$ limit stays constant at a larger temperature. The growth is more evident at high momentum modes.

\subsection{Diquark condensed phase: schematic setup}
\label{sec:diquark_condensed}

\begin{figure}[htbp]
\centering
\includegraphics*[scale=0.8]{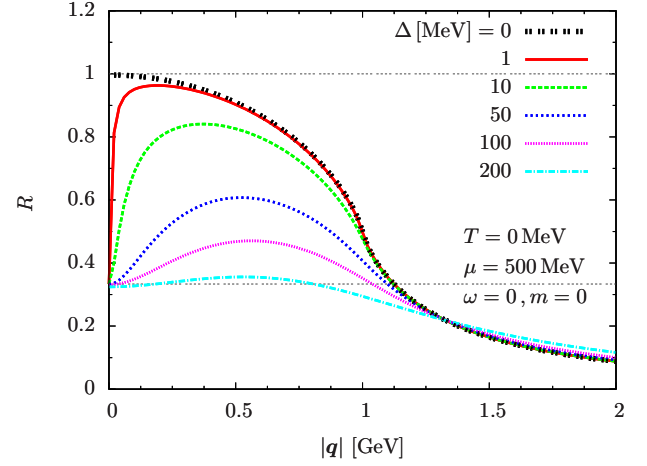}
\caption{ The $\Delta$ dependence of the CS conductivity. We set $T=m = \omega=0$ and $\mu=500$ MeV. The values of $\Delta$ are $0, 1, 10, 50, 100,$ and $200$ MeV. We show $R=1/3$ and 1 for eye-guides.
}
\label{fig:Mq00_mu05_T001_mom_dep}
\end{figure}

\begin{figure}[htbp]
\centering
\vspace{-0.5cm}
\includegraphics*[scale=0.8]{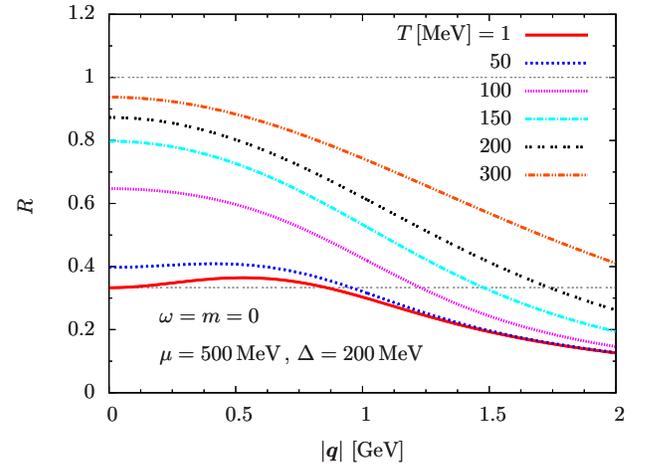}
\caption{ The $T$ dependence of the CS conductivity. We set $m = \omega=0$, $\mu=500$ MeV, and $\Delta=200$ MeV. The values of $T$ are $1, 50, 100, 150$, and $200$ MeV. 
}
\vspace{-0.2cm}
\label{fig:Mq00_del02_mu05_T_dep}
\end{figure}

Now we turn on the diquark gap. 
The major roles of $\Delta$ is to distinguish the regimes, $|{\bm q}| < \Delta $ and $ |{\bm q}| > \Delta$, and to determine the abundance of thermal quarks which is controlled by a thermal factor $ \tilde{f} \sim {\rm e}^{-\Delta/T}$. 
Its impact is substantial only for the low momentum or low temperature behaviors, $|{\bm q}| < \Delta $ or $T < \Delta$. In the other domains the results are similar to the normal phase.

Shown in Fig.\ref{fig:Mq00_mu05_T001_mom_dep} is the $\Delta$ dependence of the static CS conductivity as a function of spatial momenta ${\bm q}$. 
The values of $\Delta$ are $0, 1, 10, 50, 100, 150,$ and $200$ MeV. We set $T=m=\omega=0$ and $\mu=500$ MeV. It is clear that the size of $\Delta$ determines the domain of ${\bm q}$ where the CS conductivity deviates from that in the normal phase. With a finite $\Delta$, the static conductivity is $R=1/3$, reflecting the suppression of the particle-hole contributions by diquark gaps.

The temperature dependence of the CS conductivity can be also understood by the suppression of particle-holes. Shown in Fig.\ref{fig:Mq00_del02_mu05_T_dep} are the temperature dependence of the static CS conductivity at fixed gap, $\Delta=200$ MeV (more realistic treatments of $\Delta$ will be discussed in the next section). At low temperature $R=1/3$. As temperature increases from $T=0$ to $T\simeq \Delta$, thermal quarks contribute as in the normal phase, and $R$ increases from $R=1/3$ to the value $R= 1$ of the normal phase.

\subsection{Diquark condensed phase: realistic setup}
\label{sec:realistic_setup}

\begin{figure}[htbp]
\centering
\includegraphics*[scale=0.8]{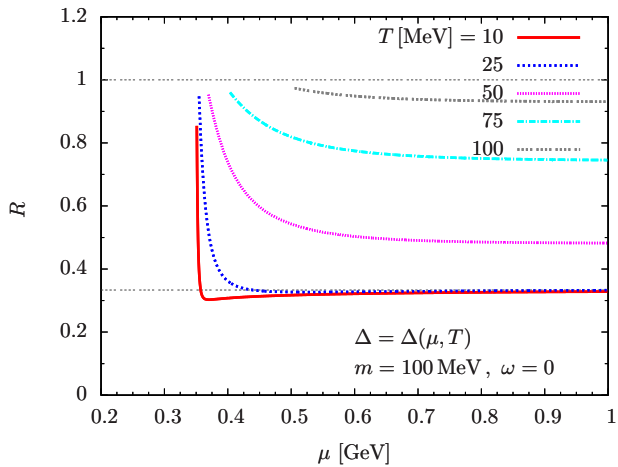}
\caption{ The $T$ and $\mu$ dependence of the static CS conductivity with $\Delta$ prepared for comparisons with the lattice simulations. We terminate the lines for low $\mu$ as our descriptions based on quarks are valid only for $\mu \ge \mu_c$ and $T<T_{\rm SF}$. The low density and temperature domain should be calculated with hadronic degrees of freedom.
}
\label{fig:lattice_setup_phase_diagram}
\end{figure}

Finally we consider a parameter set consistent with lattice simulations for QC$_2$D, and 
consider the static CS conductivity at $\omega=0$ and ${\bm q} \rightarrow 0$ for various $\mu$ and $T$. 
Most of lattice simulations have been done for relatively heavy pion mass $\simeq 700$ MeV and $\rho$ meson mass of $\sim 900$ MeV.
The onset chemical potential $\mu_c$ for the baryon density is $\mu_c = m_\pi/2 \simeq 350$ MeV.
For comparisons of our analytic results with the future lattice simulations in the diquark condensed phase, 
we use a relatively large current quark mass of $m=100$ MeV as in our previous works \cite{Kojo:2021knn,Suenaga:2019jjv}, and assume $m_\pi=700$ MeV in this setup.

Also, at this stage we introduce the $\mu$ and $T$ dependence of the gap.
For the $\mu$ dependence, we assume the zero temperature gap of the form \cite{Kogut:2000ek}
%
\beq
\Delta_0 (\mu) = \bar{\Delta}_0 \sqrt{1-(\mu_c/\mu)^4 \,} \,\theta(\mu - \mu_c)  \,,
\eeq
as predicted by the chiral effective theories.
The $\Delta_0$ vanishes at $\mu=\mu_c $, while approaches $\bar{\Delta}_0$ at high density.
 As for the temperature dependence, we assume the Bardeen-Cooper-Schrieffer (BCS) formulas as our baseline \cite{schrieffer1999theory}
\beq
T_{\rm SF} (\mu)  \simeq 0.57 \Delta_0 (\mu) \,,
\eeq
where $T_{\rm SF} (\mu)$ is the critical temperature of the diquark condensed phase at given $\mu$. The gap depends on $T$ as
\beq
\Delta (\mu,T) \simeq \Delta_{0} (\mu)\, \big( 1-T/T_{\rm SF} \big)^{1/2} \,.
\eeq
For the high density value of the diquark gap, we take $\bar{\Delta}_0 = 200$ MeV which, according to the BCS formulas, leads to $T_{\rm SF} \simeq 114$ MeV. 
This value is consistent with the lattice results $T_{\rm SF} =90-120$ MeV in the BCS domain of QC$_2$D matter for $\mu\gtrsim 500$ MeV, where the lattice results showed that $T_{\rm SF}$ depends on $\mu$ weakly. 
See Ref.\cite{Kojo:2021knn} for more detailed considerations on the applicability of the formulas.

Shown in Fig.\ref{fig:lattice_setup_phase_diagram} is the static CS conductivity for various $T$ and $\mu$. We display results only for $\mu \ge \mu_c$ and $T<T_{\rm SF}$; at lower $\mu$ and $T$ we should use the hadronic degrees of freedom. 
The normalized CS conductivity $R$ is close to $1/3$ at low temperature and approach $R=1$ at higher temperature as expected for the normal phase. The growth of $R$ is much quicker than in Fig.\ref{fig:Mq00_del02_mu05_T_dep} because now we are including the temperature dependence of gaps. For the same reason, the rapid change in $R$ at low $\mu$ happens because the $T_{\rm SF}$ is low, accordingly $\Delta$ is smaller and thermal quarks can more easily contribute to the conductivity.

We emphasize that our results on the temperature dependence are based on the assumption that thermal quarks behave as in an ideal gas of (quasi)particles; that is, once particles and holes are released from diquark condensates, they behave individually. But the validity of this picture is not immediately obvious except at very high density. The thermally excited particles and holes may need to form hadronic objects to avoid the energetic cost of the color electric flux from isolated quarks. If this is the case, the thermal factor necessary for excitations are $\sim {\rm e}^{- 2\Delta/T}$ of a particle-hole pair, instead of $\sim {\rm e}^{-\Delta/T}$ for an isolated quark \cite{Kojo:2021knn}; the thermal corrections would be much smaller than we computed.
In this respect the lattice simulations in diquark condensed phase at low density are important to delineate the properties of thermal excitations in dense matter.

\section{Discussions}
\label{sec:Discussions}

Our calculations of the CS conductivity rely on the validity of quasiparticle pictures in the diquark condensed phase. While our results on the normal phase lead to the static CS conductivity at ${\bm q}\rightarrow 0$ limit as expected from the anomaly arguments, our calculations for the diquark condensed phase lead to only the one-third; the coefficient of the CSE is not universal even in chiral limit.


This conclusion puzzled us. The triangle graph for the A-V-V type (A: axial-vector, V: vector) arises when we expand the A-V correlator by the term $\mu \gamma_0$ in the quark propagators as $S\simeq S_0 + S_0 (- \mu \gamma_0 ) S_0 + \cdots$ where $S_0$ is the propagator at $\mu=0$. However, it turned out that such expansion does not yield the same result as the usual triangle calculations.
In fact, the static CS conductivity at vanishing temperature in the diquark condensed phase can be expanded with respect to $\mu$ as
\begin{eqnarray}
\sigma_{\rm CSE} \big|^{\omega=0}_{ {\bm q} \rightarrow 0 }  = \frac{N_c}{6\pi^2}\frac{|\Delta|^2}{m^2+|\Delta|^2}\mu + {\cal O}(\mu^2)\ \ \ {\rm at}\ T=0\ . \label{SigmaMuSeries}
\end{eqnarray}
Here we have checked that the higher-order contributions ${\cal O}(\mu^2)$ are proportional to $|\Delta|^2m^2$. In this way, the conductivity is shown to be given by $S_0(-\mu\gamma_0)S_0$ part solely in chiral limit ($m\to0$), resulting in
\begin{eqnarray}
\sigma_{\rm CSE} \big|^{\omega=0}_{ {\bm q} \rightarrow 0 }  \to \frac{ C_{\rm uiv.}}{3}\mu\ , \label{SigmaUniv}
\end{eqnarray}
with $C_{\rm univ.} = N_c/(2\pi^2)$ the universal value. It should be noted that, in Eq.~(\ref{SigmaMuSeries}), the dimensionless expansion parameter is found to be $\mu/\sqrt{|\Delta|^2+m^2}$, hence $m\to0$ limit to obtain~(\ref{SigmaUniv}) has been safely taken due to a regulator played by $\Delta$. On the other hand, the other limit $\Delta\to0$ in Eq.~(\ref{SigmaMuSeries}) leads to an incorrect answer: $\sigma_{\rm CSE} \big|^{\omega=0}_{ {\bm q} \rightarrow 0 }  \to 0$, showing the diquark gap $\Delta$ changes the analytic properties of the conductivity in the infrared.

The above violation of the universality does not seems to be unnatural to us. The $\mu \gamma_0$ vertex in $S_0 (- \mu \gamma_0 ) S_0 $ introduces no momenta; the only available momentum, after integrating out the loop momenta, is the external momentum $q$. Thus the possible form from the expansion of the A-V correlator in obtaining Eq.~(\ref{SigmaUniv}) should be
\beq
\langle j^i_5 (q) j_{em}^{j} (-q) j_B^0 (0) \rangle^{\mu=0}_{A=0} ~ \mu A_{j} \sim \epsilon^{0ijk} \mu q^j A^k \,.
\eeq
%
Contracting with the external momentum $q^i$, the spatial divergence of axial current vanishes, showing that our results do not affect $\partial_\mu j^\mu_5$ and hence the anomaly relation.
From this viewpoint it is not clear to us why the CS coefficient should be constrained by the anomaly.
The possible exceptions for this discussion are the cases when $A^k$ is singular \cite{Son:2004tq,Son:2007ny}, as in the Dirac string on which $(\partial_i \partial_j - \partial_j \partial_i) {\bm A} \neq 0$, but this sort of external fields are presumably not compatible with the linear response regime utilized in this paper. 
The radiative corrections to chiral transport coefficients have been discussed in Refs.\cite{Jensen:2012jy,Gorbar:2013upa,Feng:2018tpb}.

On the other hand, our one-loop calculations shown in the previous sectors have not taken into account the vertex correction, and hence would miss some qualitatively important effects. In particular, we must treat the quasiparticle propagators and the vertices consistently in order to keep the conservation laws. Our quasiparticle propagators include diquark mean fields which break the $U(1)_A$ axial, baryon number, and electric charge conservations. The improved vertices produce the poles of NG modes that recover the conservation laws.

Below we give a brief discussion on the structure of the improved vertices and will conclude that they do not influence the conclusions in the simplest quasiparticle computations.

The vertex corrections appear both in the axial-vector current and the vector current coupled to background electromagnetic fields.
The WTI for the axial-vector vertex leads to
\beq
q^\mu \bar{ {\bf \Gamma} }_\mu^A - 2  m \Gamma_5 
= i S^{-1} (p+q) \gamma_5 + \gamma_5 i S^{-1} (p) \,,
\eeq
where the improved vertex is the sum of the bare and the correction, $\bar{ {\bf \Gamma} }_\mu^A = ( \Gamma_{\rm bare} )_\mu^5 + \delta \Gamma_\mu^A$,
with 
\beq
q^\mu \delta \Gamma_\mu^A
= 2  \left(
\begin{array}{cc}
\, 0 &  \bar{ {\bm \Delta} } \\
\, {\bm \Delta} & 0 \\
\end{array}
\right) 
= Z_A\,.
\label{WTI_A}
\eeq
[Reminder: ${\bm \Delta} =  \gamma_5\sigma^2\tau^2\Delta$ and $\bar{\bf \Delta} = -  \gamma_5\sigma^2\tau^2\Delta^* $.]
Similarly, the WTI with respect to the $U(1)_{\rm em}$ gauge symmetry leads to
\beq
q^\mu \bar{ {\bf \Gamma } }^V_\mu &=& {\cal Q}i{S}^{-1}(p+q)-i{S}^{-1}(p) {\cal Q} \,,
\eeq
where ${\cal Q}={\rm diag}(Q,-Q)$ and $\bar{ {\bf \Gamma } }^V_\mu = \Gamma^{\rm bare}_\mu + \delta\Gamma^V_\mu$ with 
\beq
q^\mu  \delta\Gamma_\mu^V 
=\left(
\begin{array}{cc}
0& Q\bar{\bm \Delta} + \bar{\bm \Delta}Q \\
-Q{\bm \Delta} - {\bm \Delta}Q & 0 \\
\end{array} 
\right) = Z_V  \,.
\label{WTI_V}
\end{eqnarray}
For Eqs.(\ref{WTI_A}) and (\ref{WTI_V}) to be valid for $q\rightarrow 0$, the correction of the vertices ($\delta \Gamma_\mu^A, \delta \Gamma_\mu^V$) must contains poles; 
otherwise the LHS would vanish in spite of the finite RHS.  
The general form of solutions is (assuming the linear dispersion between $q_0$ and ${\bm q}$)
\beq
\delta \Gamma_\mu^{s} = \frac{\, \delta^{\mu 0} q_0 + v_s \delta^{i\mu} q_i  \,}{\, q_0^2-v^2_{s} {\bm q}^{2} \,} \, Z_{s} \,,
\label{poles}
\eeq
where $s=V,A$ and $v_s$ is the medium velocity of the NG mode.
From this expression it is clear the corrections to the vertices are proportional to the external momenta ${\bm q}$. Therefore it yields a term of the form $\sim \epsilon^{0ijk} q_i q_j A_k$, and drop off from the evaluation of $\langle j_5^i \rangle$ for a regular $A_k$. Thus, while the NG modes join the anomalous processes, they do not contribute to the coefficients of the anomaly relations.

With these discussions, it seems plausible to us that the CS conductivity in chiral limit depends on the phase structure. We note that our calculations correspond to the bulk part of matter. If the coefficient of the CSE is indeed universal, it likely requires the manifest treatments of the boundaries of matter which in turn introduce space variation of the chemical potential. We expect that such boundaries accommodate surface zero modes that contribute to the CS {\it conductance} as a global quantity, rather than the conductivity as a local quantity.

\section{Conclusions}
\label{sec:Conclusions}

In this paper we have delineated the CS conductivity within the quasiparticle picture. 
The CS conductivity has been calculated in lattice simulations but the result was for the domain other than the diquark condensed phase~\cite{Buividovich:2020gnl}. We hope our considerations in this paper to be tested in near future.

The results depend on the particle-hole contributions which are sensitive to the phase structure of QC$_2$D. The particle-hole contributions are suppressed in the presence of diquark condensates. As a consequence, the static CS conductivity at low momentum limit leads to only the one-third of the normal phase. We note that this particle-hole suppression is also the origin of the electromagnetic Meissner mass, relating the CS conductivity to the Meissner effects. As temperature increases, the particles-holes come back to enhance the CS conductivity to the size of the normal phase.

In more general context, the nature of thermal corrections carry important information on the properties of matter other than the diquark condensates. Thermal excitations on top of the diquark condensed Fermi surface may be hadronic. This sort of picture has been suggested in the quarkyonic matter conjecture where the quark matter has the baryonic Fermi surface \cite{McLerran:2007qj,Kojo:2009ha,Kojo:2010fe,Kojo:2011fh,Kojo:2011cn}. We studied this point of view by examining gluon propagators in the diquark condensed phase \cite{Kojo:2021knn}, and found indications that thermal quarks, rather than thermal hadrons, induce too strong thermal corrections to gluons.
In order to derive definite conclusions, however, we need more lattice data points at low temperature and should also examine the systematics in our calculations. The CS conductivity may provide us information complementary with those from gluon propagators.

In the astrophysical aspect, the understandings of excitations in dense QC$_2$D have important applications to the physics of neutron stars \cite{Kojo:2020krb}. In this context, recently the picture of quark-hadron continuity is actively discussed \cite{Baym:2017whm,Kojo:2014rca,Kojo:2015fua,Baym:2019iky,Zhao:2020dvu,Fukushima:2020cmk,Jeong:2019lhv,Duarte:2020xsp,Duarte:2020kvi,Ma:2019ery,Ma:2021zev}) to account for the interplay between the low density nuclear physics \cite{Tews:2018kmu}, the neutron star radii for 1.4- \cite{Miller:2019cac,Riley:2019yda} and 2-solar mass ($M_\odot$) neutron stars \cite{NICER21,Miller:2021qha,Riley:2021pdl,Raaijmakers:2021uju}, and the maximum mass $\simeq 2.08\pm 0.07M_\odot$ \cite{Fonseca:2021wxt}. In particular the latest result by the NICER \cite{NICER21,Miller:2021qha,Riley:2021pdl,Raaijmakers:2021uju} shows that the radii of $1.4M_\odot$ and $2M_\odot$ neutron stars are close ($12-13$ km for both), disfavoring strong first order transitions from the nuclear to quark matter domain, and implying that there should be rather stiffening with the sound velocity exceeding the conformal 
value\footnote{To the best of our knowledge, this behavior first appeared in Ref.\cite{Masuda:2012kf,Masuda:2012ed} in the context of quark-hadron crossover model. A more general argument based on nuclear physics and neutron star observations was given in Ref.\cite{Bedaque:2014sqa}, and a microscopic description was proposed in Ref.\cite{McLerran:2018hbz} in the quakyonic matter context.
}, $\sqrt{1/3}$ \cite{Baym:2017whm,Masuda:2012kf,Masuda:2012ed,Bedaque:2014sqa,McLerran:2018hbz,Hippert:2021gfs}. 
These continuity discussions are basically for neutron star matter at zero temperature. In order to expand them into the level of finite temperature and general charge chemical potentials, we need to have more detailed insights on the excitations in dense QCD \cite{Kojo:2020ztt}.
QC$_2$D is an ideal laboratory to delineate these issues and further studies are called for.

\section*{Acknowledgement}

D. S. wishes to thank Naoki Yamamoto for useful comments. Also, the authors thank Yoshimasa Hidaka and Noriyuki Sogabe for fruitful discussions and comments.  T. K. is supported by NSFC grant No. 11875144.


\bibliography{reference}

\end{document}